\magnification=\magstep1
\tolerance=500
\vskip 2true cm
\rightline{TAUP 2753-03}
\rightline{25 November, 2003}
\bigskip
\centerline{\bf Could the Classical Relativistic Electron}
\centerline{\bf be a }
\centerline{\bf Strange Attractor?}
\bigskip
\centerline{ L.P. Horwitz${}^{a,b}$, N. Katz${}^a$, and O. Oron${}^b$}
\smallskip
\centerline{${}^a$ Department of
Physics}
\centerline{ Bar Ilan University,  Ramat Gan 529000, Israel}
\centerline{${}^b$ School of Physics and Astronomy}
\centerline{Raymond and Beverly Sackler Faculty of Exact Sciences}
\centerline{Tel Aviv University, Ramat Aviv 69978, Israel}
\bigskip
\noindent {\it Abstract:\/} We review the formulation of the problem
of the electromagnetic self-interaction of a relativistic charged
particle in the framework of the manifestly covariant classical
mechanics of Stueckeleberg, Horwitz and Piron.  The gauge fields of
this theory, in general, cause the mass of the particle to change. We
show that the non-linear Lorentz force equation for the
self-interaction resulting from the expansion of the Green's function
has chaotic solutions.  We
study the autonomous equation for the off-shell particle mass here,
 for which the effective charged particle mass
achieves a macroscopic average value determined by what appears to be
a strange attractor. 
\vfill
\break
\noindent
{\bf 1. Introduction}
\smallskip
\par The advent of powerful computers in the second half of the 
twentieth century has made it a possible and attractive possibility to
investigate physical problems that involve a high level of intrinsic
nonlinearity. The remarkable discovery by Lorenz in 1963[1] that rather
simple nonlinear systems are capable of displaying highly complex,
unstable, and, in many cases, beautiful, systems of orbits, gave rise to
the subject of chaos, now under intensive study both in specific
applications and in its general properties. The instability of these
orbits is characterized by the fact that small variations of initial
conditions generate new orbits that diverge from the original orbit
exponentially. 
\par Such studies range from
nonlinear dynamical systems, both dissipative and non-dissipative,
involving deterministic chaos, and hard surface type collisions (for
example, billliards), to the study of fluctation phenomena in quantum
field theory [2].  The results have led to a deeper
understanding of turbulence in hydrodynamics, the behavior of
nonlinear electrical circuits, plasmas, biological systems 
 and chemical reactions.  It
has, furthermore, given deep insights into the foundations of
statistical mechanics. 
\par What appears to be a truly striking fact that has emerged from
this experience, is that the appearance of what has become to be known
as chaotic behavior is not just the property of some special systems
designed for the purpose of achieving some result depending on the
nonlinearity, such as components of an electric circuit, but seems to
occur almost universally in our perception of the physical world. The
potentialities presented by this subject are therefore very extensive,
and provide a domain for discovery that appears to be virtually
unlimited. 

\par As an example of the occurrence of chaotic behavior in one of the
most fundamental and elementary systems in nature, we wish to discuss
here the results of our study of the classical relativistic charged
particle (for example, the classical electron without taking into account the
dynamical effects of its spin, an intrinsically quantum effect).
\par There remains, in the area of research on chaotic systems, 
 the question of making a 
correspondence between classical chaotic behavior and the properties
of the corresponding quantum systems.  Much is known on the signatures
of chaos for quantum systems, for example, the occurrence of Wigner
distributions in the spectra for Hamiltonian chaos.  The study of
relativistic phenomena, of the type we shall consider here, where the
explicit self-interaction problem gives rise to instability, may have
a counterpart in the renormalization program in quantum field
theory. 
 The existence of radiation due to the
accelerated motion of the particle raises the question of how this
radiation, acting back on the particle, affects its motion.  Rohrlich
[3] has described the historical development of this problem, where the first steps 
were taken by Abraham in 1905[4],
culminating in the work of Dirac[5], who derived the equation for the
ideal point electron in the form
$$ m {d^2 x^\mu \over ds^2} = F^\mu_\nu {dx^\mu \over ds} +
\Gamma^\mu, \eqno(1.1)$$
where $m$ is the electron mass, including electromagnetic correction,
 $s$ is the proper time
along the trajectory $x^\mu(s)$ in spacetime, $F^\mu_\nu$ is the
covariant form of the electromagnetic force tensor, $e$ is the
electron charge, and 
$$ \Gamma^\mu = {2 \over 3}{e^2 \over c^3} \bigl( {d^3 x^\mu \over
ds^3} - {d^2x^\nu \over ds^2}{d^2 x_\nu \over ds^2} {dx^\mu \over ds}
\bigl)
\eqno(1.2)$$
\par Here, the indices $\mu, \nu$, running over $0,1,2,3$, label the
spacetime variables that represent the action of the Lorentz group;
 the index raising and lowering Lorentz invariant tensor $\eta_\mu\nu$ is of
 the form $diag(-1,+1,+1,+1)$.  The expression for $\Gamma^\mu$ was
 originally found by Abraham in 1905[4], shortly after the discovery of
 special relativity, and is known as the Abraham four-vector of
 radiation reaction. Dirac's derivation [5] was based on a direct
 application of the Green's functions for the Maxwell fields,
 obtaining the form $(1.1)$, which we shall call the
 Abraham-Lorentz-Dirac equation.  In this calculation, Dirac
 used the difference between retarded and advanced Green's functions,
 so as to eliminate the singularity carried by each.  Sokolov and
 Terner [6], for example, give a derivation using the retarded Green's
 function alone, and show how the singular term can be absorbed into
 the mass $m$.
\par The formula $(1.1)$ contains a so-called singular perturbation
 problem.  There is a small coefficient multiplying a derivative of
 higher order than that of the unperturbed problem; since the highest
 order derivative is the most important in the equation, one sees
 that, dividing by $e^2$, any small deviation in the lower order
 terms results in a large effect on the orbit, and there are unstable
 solutions, often called ``runaway solutions''.  There is a large
 literature [7]
on the methods of treating this instability, and there is
 considerable discussion in Rohrlich's book [3] as well (he describes a
 method for eliminating the unstable solutions by studying 
 asymptotic properties).
\par The existence of these ``runaway solutions'', for which the
 electron undergoes an exponential accelaration with no external force
 beyond a short initial perturbation of the free motion, is a difficulty
 for the point electron picture in the framework of the Maxwell theory
 with the covariant Lorentz force.  Rohrlich [8] has discussed the
 idea that the point electron idealization may not be really physical,
 based on arguments from classical and 
 quantum theory, and emphasized that, for the corresponding classical 
problem, a finite size can eliminate this instability. Of course, this
 argument is valid, but it leaves open the question of
 the consistency of the Maxwell-Lorentz theory which admits the
 concept of point charges, as well as what is the nature of the
 elementary charges that make up the extended distribution.     
\par  It is quite remarkable that
Gupta and Padmanabhan[9], using essentially
geometrical arguments (solving the static problem in the frame of the
 accelerating particle with
  a curved background metric), have shown that the description of
  the motion of an accelerating charged particle {\it must} include the
 radiation terms of the Abraham-Lorentz-Dirac equation. Recognizing
that the electron's acceleration precludes the use of a sequence of
``instantaneous'' inertial frames to describe the action of the
forces on the electron [10], they carry out a Fermi-Walker
transformation [11], going
to an accelerating frame (assuming constant acceleration) in which the
electron is actually inertial, and there solve the Coulomb problem in
the curved coordinates provided by the Fermi-Walker transformation.
Transforming back to laboratory coordinates, they find the
Abraham-Lorentz-Dirac equation without the direct use of the Maxwell
Green's functions for the radiation field. This result, suggesting the
relevance of curvature in the spacetime manifold, such as that
generated by sources in general relativity, along with other,
more elementary manifestations of mass renormalization (such as the
contribution to the mass due to electromagnetic interactions and the
identification of the Green's function singularity contribution with
part of the electron mass), carries an implication that the electron mass        may
play an important dynamical role.
\par Stueckelberg, in 1941[12], proposed a manifestly covariant form of
classical and quantum mechanics in which space and time become
dynamical observables.  They are therefore represented in quantum
theory by operators on a Hilbert space on square integrable functions
in space and time.  The dynamical development of the state is
controlled by an invariant parameter $\tau$, which one might call the
world time, coinciding with the time on the (on mass shell)freely
falling clocks of general relativity.  Stueckelberg [12] started his
analysis by considering a classical world-line, and argued that under
the action of forces, the world line would not be straight, and in
fact could be curved back in time.  He identified the branch of the
curve running backward in time with the antiparticle, a view taken
also by Feynman in his perturbative formulation of quantum
electrodynamics in 1948 [11].  Realizing that such a curve could not be
parametrized by $t$ (for some values of $t$ there are two values of
the space variables), Stueckelberg introduced the parameter $\tau$ along the
trajectory. 
\par This parameter is not necessarily identical to proper
time, even for inertial motion for which proper time is a meaningful
concept. Stueckelberg postulated the existence of an invariant
``Hamiltonian'' $K$, which would generate Hamilton equations for the
canonical variables $x^\mu$ and $p^\mu$ of the form
$$ {\dot x}^\mu = {\partial K \over \partial p_\mu} \eqno(1.3)$$
and 
$$ {\dot p}^\mu =  -{\partial K \over \partial x_\mu}, \eqno(1.4)$$
where the dot indicates differentiation with respect to $\tau$.
Taking, for example, the model
$$ K_0 = {p^\mu p_\mu \over 2M}, \eqno(1.5)$$
we see that the Hamilton equations imply that
$$ {\dot x}^\mu = {p^\mu \over M} \eqno(1.6)$$
It then follows that
$$ {d{\bf x} \over dt} = {{\bf p} \over E}, \eqno(1.7)$$
where $p^0 \equiv E$, where we set the velocity of light $c=1$; this is the 
correct definition for the velocity of a free relativistic particle.
It follows, moreover, that
$$ {\dot x}^\mu  {\dot x}_\mu = {dx^\mu dx_\mu \over d\tau^2} = {p^\mu p_\mu 
\over M^2} \eqno(1.8)$$
With our choice of metric, $dx^\mu dx_\mu = -ds^2$, and $p^\mu p_\mu = -m^2$,
where $m$ is the classical experimentally measured mass of the
particle (at a given instant of $\tau$).  We 
see from this that 
$$ {ds^2 \over d\tau^2} = {m^2 \over M^2}, \eqno(1.9)$$
and hence the proper time is not identical to the evolution parameter
$\tau$.
 In 
the case that $m^2 = M^2$, it follows that  $ds = d\tau$, and we say that the 
particle is ``on shell''.
\par For example, in the case of an external potential $V(x)$, where
we write
 $x \equiv x^\mu$, the Hamiltonian becomes
$$ K = {p^\mu p_\mu \over 2M} + V(x) \eqno(1.10)$$
 so that, since $K$ is a constant of the motion, $m^2$ varies from point to point 
with the variations of $V(x)$. It is important to recognize from this discussion 
that the observable particle mass depends on the {\it state} of the system (in the 
quantum theory, the expectation value of the operator $p^\mu p_\mu$
provides
 the 
expected value of the mass squared).
 \par One may see, alternatively, that phenomenologically the mass of
a nucleon, 
such as the neutron, clearly depends on the state of the system.  The free 
neutron is not stable, but decays spontaneously into a proton, electron and 
antineutrino, since it is heavier than the proton.  However, bound in
a nucleus, 
it may be stable (in the nucleus, the proton may decay into neutron,
 positron and 
neutrino, since the proton may be sufficiently heavier than the neutron).  The 
mass of the bound electron (in interaction with the electromagnetic field), as 
computed in quantum electrodynamics, is different from that of the free electron, 
and the difference contributes to the Lamb shift.  This implies that, if one 
wishes to construct a covariant quantum theory, the variables $E$ (energy) and 
$p^\mu$ should be independent, and not constrained by the relation $E^2 = {\bf 
p}^2 + m^2,$ where $m$ is a fixed constant.  This relation
implies, moreover, that $m^2$ is a dynamical variable.  It then 
follows, quantum mechanically,  through the Fourier relation between the 
energy-momentum representation of a wave function and the spacetime 
representation, that the variable $t$, along with the variable ${\bf x}$ is a 
dynamical variable. Classically, $t$ and ${\bf x}$ are recognized as
 variables of 
the phase space through the Hamilton equations.
\par Since, in nature, particles appear
with fairly sharp mass values (not necessarily with zero spread), we may
assume the existence of some mechanism which will drive  the
particle's mass back to its original mass-shell value (after the
source responsible for the mass change ceases to act) so that the
particle's mass shell is defined.   We shall not take such a
mechanism into account explicitly here in developing the dynamical
equations. We shall assume that if this mechanism is working, it is a
relatively smooth function (for example, a minimum in free energy
which is broad enough for our off-shell driving force to work
 fairly freely)\footnote{${}^1$}{
  A relativistic
Lee model has been worked out which describes a physical mass shell
as a resonance, and therefore a stability point on the spectrum [14], but at this
point it is not clear to us how this mechanism works in general.  It
has been suggested by T. Jordan [personal communication] that the
definition of the physical mass shell could follow from the
interaction of the particle with fields (a type of
``self-interaction''); this mechanism could provide for perhaps more
than one mass state for a particle, such as the electron and muon and
the various types of neutrinos, but no detailed model has been so far
studied.}
\par In an application of statistical mechanics to this theory[15], it has
been found that a high temperature phase transition can be responsible
for the restriction of the particle's mass (on the average, in
equilibrium).  In the present work, we shall see that, at least in the
classical theory, the non-linear equations induced by radiation
reaction may have a similar effect.
\par A theory of Stueckelberg type, providing a framework for
dynamical evolution 
of a relativistic charged particle, therefore appears to be a natural dynamical
generalization of the curved space formulation of Gupta and Padmanbhan[9] and the 
static picture of Dirac.  We shall study this in the following.
\par  The 
Stueckelberg formulation implies the existence of a fifth ``electromagnetic'' 
potential, through the requirement of gauge invariance, and there is a 
generalized Lorentz force which contains a term that drives the particle 
off-shell, whereas the terms corresponding to the electric and magnetic parts of 
the usual Maxwell fields do not (for the nonrelativistic case, the electric field 
may change the energy of a charged particle, but not the magnetic field; the 
electromagnetic field tensor in our case is analogous to the magnetic
field, and the new field strengths, derived from the $\tau$ dependence
of the fields and the additional gauge field, are analogous to the
electric field, as we 
shall see).
\par In the following, we give the structure of the field equations, and show that 
the standard Maxwell theory is properly contained in this more general framework.  
Applying the Green's functions to the current source provided by the relativistic 
particle, and the generalized Lorentz force, we obtain equations of motion for the 
relativistic particle which is, in general off-shell.  As in Dirac's  result, 
these equations are of third order in the evolution paramter, and
therefore are highly unstable.  However, the equations are very
nonlinear, and give rise to chaotic behavior.
\par Our results exhibit what appears to be a strange attractor in the
phase space of the 
autonomous equation for the off-mass shell deviation, This attractor may 
stabilize the electron's mass in some neighborhood. We conjecture that
it stabilizes the orbits macroscopically as well, but a detailed
analysis awaits the application of more powerful computing facilities
 and procedures.
\bigskip
{\bf 2. Equations of motion}
\smallskip
\par The Stueckelberg-Schr\"odinger equation which governs the
evolution of a quantum state state over the manifold of spacetime was
postulated by Stueckelberg[10] to be, for the free particle,
$$ i{\partial \psi_\tau \over \partial \tau} = {p^\mu p_\mu \over
2M}\psi_\tau
\eqno(2.1)$$
where, on functions of spacetime, $p_\mu$ is represented by
$\partial/\partial x^\mu \equiv \partial_\mu$. 
\par  Taking into account
 full $U(1)$ gauge invariance, corresponding to the requirement that
 the theory maintain its form under the replacement of $\psi$ by
 $e^{ie_0 \Lambda}\psi$, 
 the Stueckelberg-Schr\"odinger 
equation (including a compensation field for the
$\tau$-derivative of $\Lambda$) is [16]
$$ \bigl(i{\partial \over \partial \tau} + e_0 a_5(x,\tau) \bigr) \psi_\tau (x) = 
  {(p^\mu - e_0a^\mu(x,\tau))(p_\mu - e_0 a_\mu(x,\tau)) \over 2M
}\psi_\tau(x), \eqno(2.2)$$
where the gauge fields may depend on $\tau$, and $e_0$ is a coupling
constant which we shall see has the dimension $\ell^{-1}$.
The corresponding classical Hamiltonian then has the form
 $$ K=  {(p^\mu - e_0a^\mu(x,\tau))(p_\mu - e_0 a_\mu(x,\tau)) \over 2M
}- e_0 a_5(x,\tau) , \eqno(2.3)$$
in place of $(2.1)$.  Stueckelberg [10] did not take into account this
full gauge invariance requirement, working in the analog of what is
known in the nonrelativistic case as the Hamilton gauge (where the
gauge function $\Lambda$ is restricted to be independent of time).
The equations of motion for the field variables are given (for both the
 classical and quantum theories) by [16]
$$ \lambda \partial_\alpha f^{\beta \alpha}(x,\tau) = e_0
j^\beta(x,\tau), \eqno(2.4)$$
where $\alpha,\, \beta = 0,1,2,3,5$, the last corresponding to the
$\tau$ index,  and $\lambda$, of dimension $\ell^{-1}$, 
 is a factor on the terms $f^{\alpha
\beta} f_{\alpha \beta}$ in the Lagrangian associated with $(2.2)$
(with, in addition, degrees of freedom of the fields) required by
 dimensionality. 
 The field strengths are 
$$ f^{\alpha \beta} = \partial^\alpha a^\beta - \partial^\beta
a^\alpha, \eqno(2.5)$$
and the current satisfies the conservation law [16]
$$ \partial_\alpha j^\alpha(x,\tau) = 0; \eqno(2.6)$$
integrating over $\tau$ on $(-\infty, \infty)$, and assuming that
$j^5(x,\tau)$ vanishes at $\vert \tau \vert \rightarrow \infty$, 
one finds that 
$$\partial_\mu J^\mu(x) = 0, $$
where (for some dimensionless $\eta$) [17] 
$$ J^\mu(x) = \eta \int_{-\infty}^\infty \, d\tau \, j^\mu(x,\tau).
\eqno(2.7)$$  
We identify this $J^\mu(x)$ with the Maxwell conserved current.
In ref. [18], for example, this expression occurs with
    $$ j^\mu (x,\tau) = {\dot x}^\mu(\tau) \delta^4(x-x(\tau)),
\eqno(2.8)$$
and $\tau$ is identified with the proper time of the particle
 (an identification which can be
made for the motion of a free particle). The conservation of the integrated current
then follows from the fact that 
$$\partial_\mu j^\mu = {\dot x}^\mu(\tau) 
\partial_\mu\delta^4(x-x(\tau))= -{d \over d\tau} \delta^4 (x-x(\tau)),$$
 a total derivative; we assume that the world line runs to infinity (at least in the
 time dimension) and therefore it's integral vanishes at the end
 points [10, 18], in accordance with the discusssion above.
 \par As for the Maxwell case, one can write the
current formally in five-dimensional form
$$ j^\alpha={\dot x}^\alpha\delta^4(x(\tau)-x). \eqno(2.9)$$ 
For $\alpha = 5$, the factor ${\dot x}^5$ is unity, and this component therefore
represents the event density in spacetime.
\par  Integrating the $\mu$-components of
Eq. $(2.4)$ over $\tau$ (assuming $f^{\mu 5} (x, \tau) \rightarrow 0$
 for $\tau \rightarrow \pm \infty$), we obtain the Maxwell
 equations with the Maxwell charge $e= e_0/\eta$ and
 the Maxwell fields  given by 
$$ A^\mu(x) = \lambda \int_{-\infty}^\infty a^\mu(x,\tau) \,d\tau. \eqno(2.10)$$
 A Hamiltonian of the form $(2.3)$ without $\tau$ dependence of the
fields, and without the $a_5$ terms, as written by Stueckelberg [10], 
 can be recovered in the limit of the zero mode of the fields (with $a_5 =0$)
in a physical state for which this limit is a good approximation
{\it i.e.}, when the Fourier transform of the fields, defined by
$$ a^\mu(x,\tau) = \int \,ds {\hat a}^\mu (x,s)
e^{-is\tau}, \eqno(2.11)$$
 has support only in the neighborhood $\Delta s$ of $s=0$.  The vector
potential then takes on the form
 $  a^\mu(x,\tau) \sim \Delta s {\hat a}^\mu(x,0)
 =(\Delta s / 2\pi \lambda) A^\mu(x)$, and we
identify $e= (\Delta s/2\pi \lambda) e_0$. The zero mode therefore
emerges when the inverse correlation length of the field $\Delta s$ is
 sufficiently small, and then  $ \eta = 2\pi \lambda/\Delta s $. We
remark that in this limit, the fifth equation obtained from $(2.4)$
decouples.  The Lorentz force obtained from this Hamiltonian, using
the Hamilton equations, coincides with the usual Lorentz force, and,
as we have seen, the generalized Maxwell equation reduce to the usual
Maxwell equations.  The theory therefore contains the usual Maxwell
Lorentz theory in the limit of the zero mode; for this reason we have
called this generalized theory the ``pre-Maxwell'' theory.
\par If such a pre-Maxwell theory really underlies the
standard Maxwell theory, then there should be some physical mechanism
which restricts most observations in the laboratory to be close to the
zero mode. For example, in a metal there is a frequency, the plasma
frequency, above which there is no transmission of electromagnetic
waves. In this case, if the physical universe is imbedded in a medium
which does not allow high ``frequencies'' to pass, the
pre-Maxwell theory reduces to the Maxwell theory. Some study has been
carried out, for a quite different purpose (of achieving a form of
analog gravity), of the properties of the
generalized fields in a medium with general dielectric tensor[19].  We
shall see in the present work that the high level of nonlinearity of
this theory in interaction with matter may itself generate an
effective reduction to Maxwell-Lorentz theory, with the high frequency
chaotic behavior providing the regularization achieved by models of
the type discussed by Rohrlich[8].
\par  We remark that
integration over $\tau$ does not bring the generalized Lorentz force
into the form of the standard Lorentz force, since it is nonlinear,
and a convolution remains.  If the resulting convolution is trivial,
i.e., in the zero mode, the two theories then coincide.  Hence, we
expect to see dynamical effects in the generalized theory which are
not present in the standard Maxwell-Lorentz theory.
\par  Writing the Hamilton equations
$${\dot x}^\mu = {dx^\mu \over d\tau} = {\partial K \over\partial p_\mu}; {\dot
p}^^\mu = {dp^\mu \over d\tau} = -{\partial K \over dx_\mu} \eqno(2.12)$$ 
 for the Hamiltonian $(2.3)$,
we find the generalized Lorentz force 
$$ M {\ddot x}^\mu = e_0 f^\mu\,_\nu {\dot x}^\nu +
f^\mu\,_5. \eqno(2.13)$$
Multiplying this equation by ${\dot x}_\mu $, one obtains
$$ M{\dot x}_\mu {\ddot x}^\mu  = e_0{\dot x}_\mu f^\mu\,_5; \eqno(2.14)$$
this equation therefore does not necessarily lead to the trivial
relation between $ds$ and $d\tau$ discussed above in
connection with Eq. $(1.9)$. The $f^\mu \,_5$ term has the effect of moving
the particle off-shell (as, in the nonrelativistic case, the energy is
altered by the electric field).
\par Let us now define
$$\varepsilon = 1 + {\dot x}^\mu {\dot x}_\mu = 1 -{ds^2 \over
d\tau^2}, \eqno(2.15)$$
where $ds^2 = dt^2 -d{\bf x}^2$ is the square of the proper
time. Since
${\dot x}^\mu = (p^\mu - e_0 a^\mu)/M$, if we interpret $p^\mu -e_0
a^\mu)(p_\mu - e_0 a_\mu) = -m^2$, the gauge invariant particle mass
[16], then
$$ \varepsilon = 1- {m^2 \over M^2} \eqno(2.17)$$
measures the deviation from ``mass shell'' (on mass shell, $ds^2 = d\tau^2$).
\bigskip
\noindent
 {\bf 3. Derivation of the differential equations for the spacetime orbit with
off-shell corrections} 
\smallskip
 \par We now review the derivation of the radiation reaction formula in the
  Stueckelberg formalism (see also [17]. Calculating the self
interaction contribution one must include the effects of the force
acting upon the particle due to its own field ($f_{self}$) in addition
to the fields generated by other electromagnetic sources
($f_{ext}$). Therefore, the generalized Lorentz force,
 using Eq.$(2.12)$ takes the form: 
$$ M{\ddot x}^\mu =  e_0{\dot x}^\nu {f_{ext}}^\mu\,_\nu+ e_0{\dot
x}^\nu {f_{self}}^\mu\,_\nu+e_0{f_{ext}}^\mu\,_5+e_0{f_{self}}^\mu\,_5
, \eqno(3.1)$$ 
  where the dynamical derivatives (dot) are with respect to the
universal
 time $\tau$, and the fields are evaluated on the event's trajectory. 
Multiplying Eq.$(3.1)$ by ${\dot x}_\mu$ we get the projected equation
$(2.14)$ in the form
$$  {M \over 2}{\dot \varepsilon} =  e_0 {\dot x}_\mu{f_{ext}}^\mu\,_5+e_0{\dot x}_\mu{f_{self}}^\mu\,_5.  \eqno(3.2)$$ 
\par The field
 generated by the current is given by the pre-Maxwell equations
 Eq.$(2.4)$,
 and choosing for it the generalized Lorentz
 gauge $\partial_\alpha a^\alpha=0$,
 we get 
$$ \lambda \partial_\alpha \partial^\alpha
 a^\beta(x,\tau)= (\sigma \partial_\tau^2 - \partial_t^2 +
 \bigtriangledown^2) a^\beta = -e_0
 j^\beta(x,\tau),  \eqno(3.3)$$ 
 where $\sigma = \pm 1$ corresponds to the possible choices of metric
 for the symmetry ${\rm O}(4,1)$ or ${\rm O}(3,2)$ of the homogeneous
 field equations.
 \par The Green's functions for Eq.$(3.3)$ can
 be constructed from the inverse Fourier transform 
$$ G(x,\tau) = { 1 \over (2\pi)^5} \int \, d^4k d\kappa {e^{i(k^\mu
 x_\mu + \sigma \kappa \tau)} \over k_\mu k^\mu   + \sigma \kappa^2}
 \eqno(3.4)$$
 Integrating this expression over all
 $\tau$ gives the Green's function for the standard Maxwell field.
 Assuming that the radiation reaction acts causally in $\tau$, we
 shall use here the $\tau$-retarded Green's function.  In his
 calculation of the radiation corrections to the Lorentz force,  Dirac
 used the difference between advanced and retarded  Green's functions
 in order to cancel the singularities that they contain.  One can,
 alternatively use the retarded Green's function and ``renormalize''
 the mass in order to eliminate the singularity [6].  In this analysis, we
 follow the latter procedure.
 \par The $\tau$- retarded Green's
 function [17] is given by multiplying the principal part
 of the integral Eq.$(3.4)$ by $\theta(\tau)$. Carrying out the
 integrations (on a complex contour in $\kappa$; we consider the case
 $\sigma = +1$ in the following), one finds (this Green's function
 differs from the $t$-retarded Green's function, constructed on a
 complex contour in $k^0$), 
$$ G(x,\tau) = {2 \theta (\tau) \over (2\pi)^3}\cases{{\tan^{-1}
\bigl({\sqrt{ -x^2-\tau^2} \over \tau} \bigr) \over (-x^2 -\tau^2)^{3
\over 2}} - {\tau \over x^2(x^2 + \tau^2)}  &$x^2 + \tau^2 < 0$; \cr
{ 1 \over 2}{1 \over (\tau^2 + x^2)^{3 \over 2}}  \ln\bigl\vert {\tau -
 \sqrt{\tau^2 + x^2} \over  \tau +
 \sqrt{\tau^2 + x^2}} \bigr\vert  - {\tau \over
x^2(\tau^2 + x^2)} &$x^2 + \tau^2 > 0$, \cr} \eqno(3.5)$$
 where we have written $x^2 \equiv x^\mu x_\mu$.     
\par With the help of this Green's function, the solutions of
   Eq.$(3.3)$
 for the self-fields (substituting the current from Eq.$(2.9)$) are
$$\eqalign{ a_{self}^\mu (x, \tau) &= {e_0 \over
   \lambda}\int\,d^4x'd\tau' G(x-x',\tau-\tau'){\dot x}^\mu(\tau')
   \delta^4(x'-x(\tau')) \cr
 &= {e_0 \over \lambda}\int \,
   d\tau' {\dot x}^\mu (\tau')G(x - x(\tau'),\tau-\tau') 
 \cr } \eqno(3.6)$$
and
$$a_{self}^5(x,\tau) 
= {e_0 \over \lambda}\int \, d\tau' G(x - x(\tau'),\tau-\tau') .
 \eqno(3.7)$$ 
\par  The Green's function is written as a scalar, acting in the same way on
 all five components of the source $j^\alpha$; to assure that the
 resulting field is in Lorentz gauge, however, it should be written as
 a five by five matrix, with the factor $\delta^\alpha_\beta -
 k^\alpha k_\beta /k^2$ ($k_5=\kappa$) included in the integrand.
 Since we  compute only the gauge invariant field strengths here,
 this extra term will not influence any of the results.
 It then follows that the generalized Lorentz force
 for the self-action (the force of the fields generated by the world
 line on a point $x^\mu(\tau)$ of
 the trajectory), along with the effect of external fields, is
 $$ \eqalign{M{\ddot x}^\mu &= {e_0^2 \over \lambda} \int\,
d\tau' ({\dot x}^\nu(\tau) {\dot x}_\nu(\tau')\partial^\mu - {\dot
x}^\nu(\tau){\dot x}^\mu(\tau') \partial_\nu)
G(x-x(\tau'))\vert_{x=x(\tau)} \cr &+ {e_0^2 \over
\lambda}\int \,d\tau' (\partial^\mu - {\dot x}^\mu(\tau')
\partial_\tau) G(x-x(\tau'))\vert_{x=x(\tau)} \cr &+ e_0
\bigl({f_{ext}}^\mu\,_\nu
 {\dot x}^\nu + {f_{ext}}^\mu\,_5 \bigr)\cr} \eqno(3.8)$$ 

 \par 
We define $ u \equiv (x_\mu(\tau) - x_\mu(\tau'))(x^\mu(\tau)
-x^\mu(\tau'))$, so that
 $$  \partial_\mu = 2(x_\mu(\tau) - x_\mu(\tau')){\partial \over
\partial u}. \eqno(3.9)$$
 Eq.$(3.8)$ then becomes 
$$\eqalign{ M{\ddot x}^\mu &= 2{e_0^2 \over \lambda} \int\,
d\tau' {\dot x}^\nu(\tau) {\dot
x}_\nu(\tau')(x^\mu(\tau)-x^\mu(\tau'))\cr &- {\dot
x}^\nu(\tau){\dot x}^\mu(\tau')(x_\nu(\tau)-x_\nu(\tau'))\}{\partial
\over \partial u}G(x-x(\tau'),\tau-\tau') \vert_{x=x(\tau)}\cr
 &+ {e_0^2 \over \lambda}\int \,d\tau'
(2(x^\mu(\tau)-x^\mu(\tau')){\partial \over \partial u}- {\dot
x}^\mu(\tau')\partial_\tau\}
G(x-x(\tau'),\tau-\tau')\vert_{x=x(\tau)} \cr  &+ e_0
\bigl({f_{ext}}^\mu\,_\nu {\dot x}^\nu
 +{f_{ext}}^\mu\,_5 \bigr). \cr} \eqno(3.10)$$ 
\par In the self-interaction problem where $\tau \rightarrow \tau'$,
$x^\mu(\tau')-x^\mu(\tau) \rightarrow 0$, the Green's function is
very divergent. Therefore one can expand all expressions in
$\tau''=\tau-\tau'$ assuming that the dominant contribution is from
the neighborhood of small $\tau''$ . The divergent terms are later
absorbed into the mass and charge definitions leading to
renormalization (effective mass and charge). Expanding the integrands in 
Taylor series around the
most singular point $\tau'=\tau$ and keeping the lowest order terms, the
variable $u$ reduces to \footnote{$^2$}{In ref. [17], we
considered $u\cong {\dot x}^\mu{\dot x}_\mu{\tau''}^2$ omitting
several possibly significant terms. Therefore the corrected equations
 here contain these additional terms.}  :
 $$u\cong {\dot x}^\mu{\dot x}_\mu{\tau''}^2-{\dot
x}_\mu{\ddot x}^\mu \tau''^3+{1\over 3}{\dot x}_\mu
{\mathop{x}^{...}}^\mu \tau''^4+{1\over
 4}{\ddot x}_\mu{\ddot x}^\mu \tau''^4.\eqno(3.11)$$

We now recall the definition of the off-shell deviation, $\varepsilon$
 given
 in Eq.$(2.15)$, along with its derivatives: 
$$\eqalign{ {\dot x}_\mu{\dot x}^\mu &= -1+\varepsilon \cr
 {\dot x}_\mu{\ddot x}^\mu &= {1\over 2}{\dot \varepsilon}\cr
 {\dot x}_\mu {\mathop{x}^{...}}^\mu &+ {\ddot x}_\mu{\ddot x}^\mu
= {1\over 2}{\ddot
 \varepsilon}.\cr}
\eqno(3.12)$$
Next, we define
 $$\eqalign{ w &\equiv {1 \over 12}  {\dot x}_\mu
{\mathop{x}^{...}}^\mu+{1 \over 8}{\ddot \varepsilon} \cr
\Delta &\equiv -{1 \over 2}
{\dot \varepsilon} \tau''+w{\tau''}^2. \cr}\eqno(3.13)$$
\par  Using these definitions along with those of
 Eq.$(3.11)$ and Eq.$(3.12)$ we find 
$$ {u+\tau''^2 \over \tau''^2}\cong \varepsilon+\Delta
\eqno()$$
 We then expand the Green's function
 to leading orders: 
$$\eqalign{ {\partial G \over
\partial u} &\cong {\theta(\tau'')f_1 (\epsilon+\Delta)\over (2\pi)^3
{\tau''}^5}={\theta(\tau'')\over (2\pi)^3}\bigl[{f_1(\varepsilon)\over
{\tau''}^5}-{{\dot \varepsilon} f'_1(\varepsilon)\over 2
{\tau''}^4}+\bigl(wf'_1(\varepsilon)+{1\over 8}{\dot
\varepsilon}^2f''_1(\varepsilon)\bigr){1\over {\tau''^3}}\bigr] \cr
 {{\partial G}\over{\partial\tau''}} &\cong {{2
\theta(\tau'')f_2 (\epsilon+\Delta)}\over {(2\pi)^3{\tau''}^4}}+ {{2
\delta(\tau'')f_3 (\epsilon+\Delta)}\over {(2\pi)^3{\tau''}^3}}=
\cr  &= \, \, {2 \theta(\tau'')\over
(2\pi)^3}\bigl[{f_2(\varepsilon)\over {\tau''}^4}-{{\dot \varepsilon}
f'_2(\varepsilon)\over 2
{\tau''}^3}+\bigl(wf'_2(\varepsilon)+{1\over8}{\dot
\varepsilon}^2f''_2(\varepsilon)\bigr){1\over {\tau''^2}}\bigr]+
\cr  &+ \, \, {2 \delta(\tau'')\over
(2\pi)^3}\bigl[{f_3(\varepsilon)\over {\tau''}^3}-{{\dot \varepsilon}
 f'_3(\varepsilon)\over 2
{\tau''}^2}+\bigl(wf'_3(\varepsilon)+{1\over 8}{\dot
\varepsilon}^2f''_3(\varepsilon)\bigr){1\over \tau''}\bigr)  
 \cr} \eqno(3.15)$$
 where $f' \equiv {df \over d \varepsilon }$ and 
 \bigskip
\noindent for $\varepsilon < 0$,

$$\eqalign{ f_1(\varepsilon) &= {3\tan^{-1}(\sqrt{-\varepsilon})
 \over{(-\varepsilon)}^{5 \over 2}}-{3 \over
 \varepsilon^2(1-\varepsilon)}+{2 \over \varepsilon(1-\varepsilon)^2}
 \cr f_2(\varepsilon) &= {3\tan^{-1}(\sqrt{-\varepsilon})
 \over{(-\varepsilon)}^{5 \over 2}}-{1\over \varepsilon^2}-{2 -
 \varepsilon  \over \varepsilon^2(1-\varepsilon)} \cr
 f_3(\varepsilon) &= {\tan^{-1}(\sqrt{-\varepsilon})
 \over{(-\varepsilon)}^{3 \over 2}}+{1 \over
 \varepsilon (1-\varepsilon)}\cr}. \eqno(3.16)$$
\bigskip
\noindent For $\varepsilon>0$,
 $$\eqalign{ f_1(\varepsilon) &= {{3\over
2}\ln\bigl\vert{1+\sqrt{\varepsilon} \over
 1-\sqrt{\varepsilon}}\bigr\vert \over{(\varepsilon)}^{5 \over 2}}-{3
 \over \varepsilon^2(1-\varepsilon)}+{2 \over
 \varepsilon(1-\varepsilon)^2}\cr  f_2(\varepsilon) &=
 {{3\over 2}\ln\bigl\vert{1+\sqrt{\varepsilon} \over
 1-\sqrt{\varepsilon}}\bigr\vert \over{(\varepsilon)}^{5 \over
 2}}-{1\over \varepsilon^2}-{2 - \varepsilon  \over
 \varepsilon^2(1-\varepsilon)}\cr  f_3(\varepsilon) &= \ -
 {{1\over 2}\ln\bigl\vert{1+\sqrt{\varepsilon} \over
 1-\sqrt{\varepsilon}}\bigr\vert \over\varepsilon^{3 \over
 2}}+{1\over
 \varepsilon(1-\varepsilon)}. \cr} \eqno(3.17)$$
\par  For either sign of $\varepsilon$, when $\varepsilon \sim 0$,
 $$\eqalign{ f_1(\varepsilon) &\sim  {8 \over 5} +{24 \over
 7}\varepsilon+{16 \over 3}{\varepsilon}^2+O(\varepsilon^3), \cr
 f_2(\varepsilon) &\sim  -{2 \over 5} - {4 \over 7}\varepsilon-{2
 \over 3}{\varepsilon }^2+O(\varepsilon^3),\cr
 f_3(\varepsilon) &\sim  {2 \over 3} + {4 \over 5}\varepsilon+{6
 \over 7}{\varepsilon}^2+O(\varepsilon^3). \cr} \eqno()$$         
One sees that the derivatives in Eq.$(3.15)$ have no singularity in
 $\varepsilon$ at $\varepsilon=0$.
\par From Eq.$(3.6)$ and Eq.$(3.7)$, we have 
$$\eqalign{ {f_{self}}^\mu \,_5 (x(\tau), \tau) &=\cr e
\int \,(2(x^\mu(\tau)&-x^\mu(\tau')){\partial\over \partial u}-
{\dot x}^\mu(\tau') \partial_\tau\} \times \cr &\times
G(x-x(\tau'),\tau-\tau')
\vert_{x=x(\tau)}d\tau'.\cr}\eqno()$$
 We then expand $x^\mu(\tau) -x^\mu(\tau')$ and ${\dot
x}^\mu(\tau) -{\dot x}^\mu(\tau')$  in power
series in $\tau''$, and write the integrals formally with infinite
limits. \par Substituting Eq.$(3.19)$ into Eq.$(3.2)$,
we obtain (note that $x^\mu$ and its derivatives are evaluated at the
point $\tau$, and are not subject to the $\tau''$ integration), after
integrating by parts using $\delta(\tau'')= {\partial\over
 \partial \tau''}\, \theta(\tau'')$,

$$\eqalign{ {M \over 2} {\dot \varepsilon} &= {2e_0^2 \over
\lambda (2\pi)^3} \int_{-\infty}^\infty d\tau''\bigl\{{g_1 \over
\tau''^4}(\varepsilon-1)- {g_2\over\tau''^3} {{\dot \varepsilon} \over
2} +  {g_3\over \tau''^2} {\dot x}_\nu {\mathop{x}^{...}}^\nu-
\cr &- {h_1 \over {\tau''}^3}(\varepsilon-1){\dot
\varepsilon}+{h_2 \over {2 \tau''}^2}{\dot \varepsilon}^2+{h_3 \over
\tau''^2}(\varepsilon-1)w+\cr &+{h_4\over 8\tau''^2}{\dot
\varepsilon}^2(\varepsilon-1) \bigr\} \theta (\tau'')
+e_0{\dot x}_\mu{f_{ext}}^\mu\,_5 , \cr} \eqno(3.20)$$
 where we have defined
 $$\eqalign{ g_1 &= f_1-f_2-3f_3 \,\, , \,\,g_2={1 \over
2}f_1-f_2-2f_3 \,\, , \,\,g_3={1 \over 6}f_1-{1 \over 2}f_2-{1 \over
2}f_3 \cr  h_1 &={1 \over 2}f'_1-{1 \over 2}f'_2-f'_3 \,\, ,
\,\,h_2={1 \over 4}f'_1-{1 \over 2}f'_2-{1 \over 2}f'_3\,\, ,
\,\,h_3=(f'_1-f'_2-f'_3) \cr
 h_4 &= f''_1-f''_2-f''_3.\cr}\eqno(3.21)$$

\par The integrals are divergent at the lower bound $\tau''=0$ imposed by
 the $\theta$-function; we therefore take these integrals to a cut-off
 $\mu >0$. 
 Eq.$(3.20)$ then becomes
 $$\eqalign{ 
{M \over 2} {\dot \varepsilon} &= {2e_0^2 \over \lambda (2\pi)^3}
 \bigl\{{g_1 \over 3\mu^3}(\varepsilon-1)- {g_2\over 4 \mu^2} {\dot
 \varepsilon} + 
 {g_3\over \mu} {\dot x}_\nu {\mathop{x}^{...}}^\nu- \cr
&- {h_1 \over 2 \mu^2}(\varepsilon-1){\dot \varepsilon}+{h_2 \over 2
\mu}
{\dot \varepsilon}^2+{h_3 \over \mu}(\varepsilon-1)w+ \cr
 &+ {h_4\over 8\mu}{\dot \varepsilon}^2(\varepsilon-1) \bigr\}
 \theta (\tau'')+e_0{\dot x}_\mu{f_{ext}}^\mu\,_5 . \cr} \eqno(3.22)$$
\par Following a similar procedure, we obtain from Eq.$(3.8)$

$$\eqalign{ M{\ddot x}^\mu &= {2e_0^2 \over \lambda (2\pi)^3}
 \bigl\{ -{1\over 2}\bigl((1-\varepsilon){\ddot x}^\mu+{{\dot
 \varepsilon}\over 2}{\dot x}^\mu \bigr) \bigl({f_1\over 2
 \mu^2}-{{\dot \varepsilon}f'_1 \over 2 \mu}\bigr) + {f_1\over 3 \mu}
 \bigl({\dot x}_\nu{\mathop{x}^{...}}^\nu {\dot
 x}^\mu+(1-\varepsilon){\mathop{x}^{...}}^\mu\bigr)\cr
 &+{g_1 \over 3 \mu^3}{\dot x}^\mu-{g_2 \over 2 \mu^2}{\ddot
 x}^\mu+{g_3 \over \mu}{\mathop{x}^{...}}^\mu-{h_1 {\dot  \varepsilon}
 \over 2 \mu^2}{\dot x}^\mu+{h_2 {\dot  \varepsilon} \over \mu}{\ddot
 x}^\mu+{h_3 w\over \mu}{\dot x}^\mu+{h_4{\dot \varepsilon}^2 \over 8
 \mu}{\dot x}^\mu \bigr\} \cr  &+ e_0
 \bigl({f_{ext}}^\mu\,_\nu {\dot x}^\nu +{f_{ext}}^\mu\,_5  \bigr).
 \cr}\eqno(3.23)$$
\smallskip 
Substituting Eq.$(3.22)$ for the coefficients of the ${\dot x}^\mu$
 terms in the second line of Eq.$(3.23)$ we find:
$$\eqalign{
 M(\varepsilon, {\dot \varepsilon}) {\ddot x}^\mu &= \,-  {1 \over 2}{M(\varepsilon) \over
 (1-\varepsilon)} {\dot \varepsilon} {\dot x}^\mu+ {2e_0^2 \over
 \lambda (2\pi)^3 \mu}  F(\varepsilon) \bigl\{{\mathop{x}^{...}}^\mu +
 { 1\over (1-\varepsilon)} {\dot x}_\nu{\mathop{x}^{...}}^\nu {\dot
 x}^\mu \bigr\} \cr &+ {e_0{\dot x}^\mu{\dot
 x}_\nu{f_{ext}}^\nu\,_5 \over 1-\varepsilon}
 +e_0{f_{ext}}^\mu\,_\nu {\dot x}^\nu +e_0{f_{ext}}^\mu\,_5  ,  
 \cr} \eqno(3.24)$$
where
$$ F(\varepsilon)={f_1 \over 3}(1-\varepsilon)+g_3.\eqno(3.25)$$ 
\par Here, the coefficients of ${\ddot x}^\mu$ have been grouped
formally into a renormalized (off-shell) mass term, defined (as done
in the standard
 radiation reaction problem ) 
$$ M(\varepsilon, {\dot \varepsilon}) = M + {e^2 \over
2\mu}\bigl[{f_1(1-\varepsilon) \over 2}+g_2\bigr]-e^2\bigl[{1\over
4}f_1'(1-\varepsilon)+h_2\bigr]{\dot
 \varepsilon}, \eqno(3.26)$$ 
where, as we shall see below, 
$$ e^2 =  {2e_0^2 \over \lambda (2\pi)^3\mu} \eqno(3.27)$$
 can be identified with the Maxwell charge by studying the on-shell
 limit.
\par We remark that one can change variables, with the help of
$(2.15)$ (here, for simplicity, assuming $\varepsilon <1$), to obtain
a differential equation in which all derivatives with respect to
$\tau$ are replaced by derivatives with respect to the proper time
$s$.  The coefficient of the second derivative of $x^\mu$ with respect
to $s$, and ``effective mass'' for the proper time equation, is then
given by
 $$\eqalign{ M_S(\varepsilon, {\dot \varepsilon})&={2 \over 3
F(\varepsilon)(1-\varepsilon)}\bigr\{M +{e^2 \over 2
c^3\mu}\bigl[{f_1(1-\varepsilon) \over 2}+g_2\bigr]\cr  &-{e^2\over
c^3}\bigl
[{1\over 4}f_1(1-\varepsilon)+h_2+{3 \over
2}F(\varepsilon)\sqrt{1-\varepsilon}\bigr]{\dot
\varepsilon}\bigr\}\cr} \eqno(3.26')$$
 \par Note that the renormalized mass depends on $\varepsilon (\tau)$; for
 this quantity to act as a mass, $\varepsilon$ must be slowly varying
 on some interval on the orbit of the evolution compared to all other
 motions. The computer analysis we give below indeed shows that there
are large intervals of almost constant $\varepsilon$. In case, as at
some points, $\varepsilon$ may be rapidly
 varying, one may consider the definition
 Eq.$(3.26)$ as formal; clearly, however, if $M(\varepsilon, {\dot
 \varepsilon})$ is large, ${\ddot x}^\mu$ will be suppressed (e.g., for
 $\varepsilon$ close to unity, where $ M_S(\varepsilon, {\dot
 \varepsilon})$ goes as ${\dot \varepsilon}/(1-\varepsilon)^2$ and 
$ M(\varepsilon, {\dot \varepsilon})$ as 
  ${\dot \varepsilon}/(1-\varepsilon)^3$; note that $F(\varepsilon)$
 goes as $(1-\varepsilon)^{-3}$). 
 \par  We now obtain, from Eq.$(3.24)$,
 $$\eqalign{ M(\varepsilon) {\ddot x}^\mu &= -{1 \over 2}
 {M(\varepsilon)\over {1-\varepsilon}}{\dot \varepsilon} {\dot x}^\mu
 + F(\varepsilon) e^2\bigl\{ {\mathop{x}^{...}}^\mu+ {1 \over
 1-\varepsilon}{\dot x}_\nu {\mathop{x}^{...}}^\nu {\dot x}^\mu
 \bigr\} \cr &+ e_0 {f_{ext}}^\mu_\nu {\dot x}^\nu +
 e_0\Bigl({{\dot x}^\mu {\dot x}_\nu \over
 1-\varepsilon}+\delta^\mu_\nu
 \Bigr){f_{ext}}^\nu_5.\cr} \eqno(3.28)$$
\par We remark that when one multiplies this equation by ${\dot x}_\mu$, it
 becomes an identity (all of the terms except for $ e_0 {f_{ext}}^\mu_
 \nu{\dot x}^\nu$ may be grouped to be proportional to $\Bigl( {{\dot
 x}^\mu{\dot x}_\nu \over 1-\varepsilon}+\delta^\mu_\nu \Bigr)$); one
 must use Eq.$(3.22)$ to compute the off-shell mass shift
 $\varepsilon$ corresponding to the longitudinal degree of freedom in
 the direction of the four velocity of the particle.  Eq.$(3.28)$
 determines the motion orthogonal to the four
 velocity. Equations $(3.22)$ and $(3.28)$ are the
 fundamental dynamical equations governing
 the off-shell orbit. 
\par We remark  that as in [17] it can
be shown that Eq.$(3.28)$ reduces to the ordinary
 (Abraham-Lorentz-Dirac)
 radiation
reaction formula for small, slowly changing $\varepsilon$ and that
that no instability, no radiation, and no acceleration of the
 electron occurs when it is on shell. There is therefore no ``runaway
 solution'' for the exact mass shell limit of this theory; the
 unstable Dirac result is approximate for $\varepsilon$ close to, but
 not precisely zero.
\bigskip  
\noindent
{\bf 4. The $\varepsilon$ evolution}
 \smallskip
 \par We now derive an equation for the evolution of
  the off-shell deviation, $\varepsilon$, when the external field is
  removed. We then use this equation to prove that a fixed mass-shell
  is consistent only if the particle is not accelerating, and
  therefore no runaway solution occurs.
  Using the definitions
$$\eqalign{ F_1(\varepsilon) &= { g_1(\varepsilon-1)\over
3\mu^2}\cr F_2(\varepsilon) &= {g_2-2(\varepsilon-1)h_1
\over 4\mu} \cr F_3(\varepsilon) &= g_3+{1 \over
12}(\varepsilon-1)h_3 \cr F_4(\varepsilon) &={1\over
2}h_2+{1\over 8}(\varepsilon-1)h_4 \cr F_5(\varepsilon)
&={1\over 8}(1-\varepsilon)h_3 \cr}\eqno(4.1)$$
in equation Eq.$(3.22)$, in the absence of external fields, we write 
$${\dot x_\mu  {\mathop{x}^{...}}^\mu}={1\over
F_3(\varepsilon)}\bigr\{{M\over 2e^2}{\dot
\varepsilon}-F_1(\varepsilon)+F_2(\varepsilon){\dot
\varepsilon}-F_4(\varepsilon){\dot
\varepsilon}^2-F_5(\varepsilon){\ddot \varepsilon}
\bigr\}. \eqno(4.2)$$
 Differentiating with respect to
 $\tau$ we find:
 $$\eqalign{ {\dot x_\mu  {\mathop{x}^{....}}^\mu}&+{\ddot x_\mu
{\mathop{x}^{...}}^\mu}={1 \over F_3}\bigl\{F'_2{\dot
\varepsilon}^2+{\ddot \varepsilon}\bigl({M\over 2 e^2}+F_2 \bigr)-F'_1
{\dot \varepsilon} -F'_4(\varepsilon){\dot
\varepsilon}^3-\bigl(2F_4(\varepsilon)+ \cr&-F'_5(\varepsilon)\bigr){\dot \varepsilon}{\ddot
\varepsilon}-F_5{\mathop{\varepsilon}^{...}}\bigr\}-{F'_3  \over
{F_3}^2}\bigl\{F_2+ {M \over 2 e^2}{\dot
\varepsilon}-F_1-F_4(\varepsilon){\dot
\varepsilon}^2-F_5(\varepsilon){\ddot \varepsilon} \bigr\}{\dot
\varepsilon} \cr&\equiv H-{F_5 \over
F_3}{\mathop{\varepsilon}^{...}}. 
 \cr}\eqno(4.3)$$

Together with 
$${\dot x_\mu  {\mathop{x}^{....}}^\mu}+3{\ddot x_\mu 
 {\mathop{x}^{...}}^\mu}= {1\over 2}{\mathop{\varepsilon}^{...}},\eqno(4.4)$$ 
which is the $\tau$ derivative of the last equation in
 Eq.$(3.11)$,
 one finds from Eq.$(4.3)$, 
$$ {\ddot x_\mu  {\mathop{x}^{...}}^\mu}=\bigl({1\over
4}+{F_5 \over2 F_3}(\varepsilon)\bigr)
{\mathop{\varepsilon}^{...}}-{1\over2}H(\varepsilon,{\dot
\varepsilon},
 {\ddot \varepsilon}). \eqno(4.5)$$ 
Multiplying Eq.$(3.28)$ by ${\ddot x}_\mu$ (with no external
 fields)
 and using Eq.$(4.2)$ and Eq.$(4.5)$, we obtain 
$$  \bigl(1+2{F_5\over
F_3}\bigr){\mathop{\varepsilon}^{...}}-A(\varepsilon){\ddot
\varepsilon}+ B(\varepsilon){\dot \varepsilon}^2 +
 C(\varepsilon) {\dot \varepsilon}-D(\varepsilon)+E(\varepsilon){\dot
\varepsilon}^3+I(\varepsilon){\dot \varepsilon}{\ddot \varepsilon}=0,
 \eqno(4.6)$$ 
where 
$$\eqalign{ A &= {2 \over F_3}\bigl( {M \over 2e^2} + F_2 \bigr)
+ {2 M(\varepsilon) \over e^2 F(\varepsilon)}-{4M(\varepsilon)F_5\over
2e^2F(\varepsilon)},\cr B&= {2F_3' \over F_3^2}\bigl(F_2-
{M\over 2e^2} \bigr) - {2F_2' \over F_3} + {2 \over {1-\varepsilon}}
{1 \over F_3} \bigl( {M\over 2e^2} + F_2 \bigr)-{M(\varepsilon)\over
e^2F(\varepsilon)} {1 \over 1-\varepsilon}, \cr C&= {4
M(\varepsilon) \over e^2 F(\varepsilon)} {1 \over F_3}\bigl( {M\over
2e^2} + F_2 \bigr) - {2  \over F_3^2}F_1 F_3'-{2F_1 \over
(1-\varepsilon)F_3} + {2 \over F_3}F_1' ,\cr D &= {4
M(\varepsilon) \over e^2 F(\varepsilon)}{F_1 \over F_3} \cr
E&={2F_4\over (1-\varepsilon)F_3}-2\bigl({F_4F'_3\over
{F_3}^2}-{F'_4\over F_3}\bigr)\cr I &= 2 \bigl({F'_5\over
F_3}+2{F_4\over F_3}-{F_5F'_3\over{F_3}^2}\bigr)-
{2F_5 \over(1-\varepsilon)F_3}.\cr} \eqno(4.7)$$     
\bigskip
\noindent
{\bf 5. Dynamical Behavior of System}
\smallskip
\par In this section, we present results of a preliminary study of the
 dynamical behavior of the system.  The four orbit equations pose a
very heavy computational
problem, which we shall treat in a later publication.  We remark
here, however, that the quantities $d{\bf x}/d\tau$ and $dt/d\tau$
appear to rise rapidly, but the ratio, corresponding to the observed
velocity $d{\bf x}/dt$ actually falls, indicating that what would be
observed is a dissipative effect. 
\par We shall concentrate here on the evolution of the off-shell mass 
correction $\varepsilon$, and show that there is a highly complex dynamical
behavior showing strong evidence of chaos.
\par It appears that $\varepsilon$ reaches values close to (but less
than) unity, and 
there exhibits very complex behavior, with large fluctuations.  The
effective mass
$M(\varepsilon, {\dot \varepsilon})$ therefore appears to reach a macroscopically steady
value, apparently a little less than unity.  In this neighborhood, it
is easy to see from the expression $(3.26)$ that the effective mass
itself becomes very large, as ${\dot \varepsilon}/(1-\varepsilon)^3$,
and therefore the effect of the Lorentz forces in producing
acceleration becomes small.  
\par The third order equation of $\varepsilon$, Eq.$(4.6)$, can be
written as
 a set of three first order equations of the form
$$ {d{\bf \gamma} \over dt} = H({\bf \gamma}), \eqno(5.1)$$
where ${\bf \gamma} = \gamma 1, \gamma 2, \gamma 3 = \varepsilon,
 {\dot \varepsilon}, {\ddot \varepsilon}$.  We have obtained
numerical 
solutions over a range of variation of $\tau$.  In figure 1, we show,
on a
 logarithmic scale, these three functions; one sees rapid fluctuations
in $\gamma 2 $ and $\gamma 3$, and large variations in $\gamma 1$.
The action of what appears to be an attractor occurs on these graphs
in the neighborhood of $\tau = 0.002$ (time steps were taken to be of
the order of $10^{-6}$). The time that the system remains on this
attractor, in our calculation about
four times this interval, covers about three cycles. The characteristics
of the orbits in phase space are very complex, reflecting an evolution
of the nature of the attractor.
\par The dynamical significance of
these functions is most clearly seen in
 the phase space plots.  In figure 2, we show $ {\ddot \varepsilon} =
 \gamma 3 $
vs. ${\dot \varepsilon} = \gamma 2$. This figure shows the approach to
the apparent attractor inducing motion beginning at the lower convex
portion of the unsymmetrical orbit which then turns back in a
characteristic way to reach the last point visible on this graph. We
then continue to examine the motion on a larger scale, where we show,
in figure 3, that this end point opens to a larger and more symmetric
pattern.  The orbit then continues, as shown in figure 4 to a third
loop,  much smaller, which then reaches a structure that is
significantly different.  The end region of this cycle is shown on
larger scale in figure 5, where a small cycle terminates in
oscillations that appear to become very rapid as $\varepsilon$
approaches unity.  The coefficients in the differential equation
become very large in this neighborhood.
\par In figure 6 we see the approach to the attractor in the graph of
${\dot
 \varepsilon} = \gamma 2 $ vs. $\varepsilon = \gamma 1$; this curve
 enters a region of folding, which develops to a loop shown in figure
 7a,and in fig. 7b, on a larger
 scale, where we had to terminate the calculation due to limitations
 of our computer.
\par We have computed the global Lyapunov exponents according to the
 standard procedures; recognizing, however, that there appear to be
 (at least) two levels of behavior types, we have studied the
 integration on phase space for $\varepsilon$ somewhat less than unity, since
 the integration diverges for $\varepsilon$ close
 to one. For  $\varepsilon$ very close to one, in fact, due to the
 relation $ds = (1-\varepsilon)d\tau$, this neighborhood contributes
 very little to the development of the observed orbits.
\par In this calculation, we have computed the largest Lyapunov
 exponent (positive) by studying the average separation of the orbits
 associated
with nearby initial conditions [20]. We show the develpment of the
 Lyapunov exponent as the average is taken over an increasing interval in
 fig. 8 as data is accumulated up to 13,800 iterations (at least
 two cycles on the attractor).   Note that there are intervals of
 relative stability and very strong instability. 
\par Starting the calulation with different initial conditions, we 
found very similar dynamical behavior, confirming the
 existence of the apparent attractor. 
\par We have also studied the time series associated with the three
 variables of the phase space, shown in figures 9-11.  The
 autocorrelation
 functions for
 ${\ddot \varepsilon}$ and ${\dot \varepsilon}$ fall off quickly
 giving characteristic correlation times; the autocorrelation function
 for $\varepsilon$ does not fall off faster than linearly, but the
 time series shows significant structure resembling the phase space
 structure (subtracting out the local average from the function
 $\varepsilon$ generates a deviation function that does indeed show a
 fall off of the autocorrelation function). We have used these scales,
 for 20 time steps (for ${\ddot \varepsilon}$ and $\varepsilon$) and 
 40 time steps (for ${\dot \varepsilon}$, to contruct a three dimensional delay
 phase space for the time series for each of these functions.
\par In figs. 12a and 12b, we show the behavior of solution of the
 autonomous equation for ${\dot \varepsilon}$ vs. $\varepsilon$ and
 ${\ddot \varepsilon}$ vs. ${\dot \varepsilon}$.  These results are
 very similar in form to those given above, demonstrating
  stability of the results. 
\bigskip
\noindent
{\bf 6. Discussion}
\smallskip
\par We have examined the equations of motion generated for the
classical relativistic charged particle, taking into account
dynamically generated variations in the mass of the particle, and the
existence of a fifth electromagnetic type potential essentially
reflecting 
the gauge degree
of freedom of the variable mass.  We have focussed our attention on
the autonomous equation for the off-shell mass deviation defined in
$(2.15)$ and $(2.17)$ as an indication of the dynamical behavior of
the system, and shown that there indeed appears to be a highly
complex strange attractor.  It is our conjecture that the formation of
dynamical attractors of this type in the orbit equations as well will
lead to a macroscopically smooth behavior of the perturbed
relativistic charged particle, possibly associated with an effective
non-zero size, as discussed, for example,  by Rohrlich[8]. Some
evidence for such a smooth behavior appears in the very large
effective mass generated near $\varepsilon \cong 1$, stabilizing the
effect of the Lorentz forces, as well as our preliminary result that
the larger accellerations emerging in derivatives with respect to
$\tau$ are strongly damped, in this neighborhood of $\varepsilon$, in
the transformation to motion seen as velocity ($d{\bf x}/dt$) in a
 particular frame.
\par Our investigation of the phase space of the autonomous equation
used somewhat arbitrary initial conditions, and we find that changing
these conditions somewhat does not affect the general pattern of the
results.  Moreover, since ${\ddot \varepsilon}$ depends on
the third derivative of $x^\mu(\tau)$, this initial condition should
not be taken as arbitrary, but as determined by the results for the orbit
(initial conditions for the orbit depend only on $x^\mu$ and its first
two derivatives).  Computing the orbit for a typical choice of initial
conditions, and using the result to fix  ${\ddot \varepsilon}$, we
find that the corresponding solution of the autonomous equation indeed
lies in the attractor.  The results of our investigation of the
 autonomous equation should therefore be valid for initial conditions
that obey the physical constraints imposed by the equations of motion
as well. 
   \par We have specified all of the dynamical equations here, but a
complete investigation must await further planned work with more powerful
computing facilities and procedures. 
\bigskip
 \noindent
{\bf Acknowledgement}
\bigskip
\par One of us (L.P.H.) would like to thank  the Institute for
Advanced Study, Princeton, N.J., for partial support, and Steve Adler
for his hospitality, during his visit in the Spring Semester (2003)
when much of this work was done.
    \bigskip
\frenchspacing
{\bf References}
\item{1.} E. Lorenz, Jour. Atmos. Sci. {\bf 20}, 130 (1963).
\item{2.} See, for example, C. Beck, {\it Spatio-Temporal Chaos and 
Vacuum Fluctuations
of Quantized Fields\/}, World Scientific, Advanced Series in Nonlinear
Dynamics, vol. 21, Singapore (2002).
\item{3.} F. Rohrlich, {\it Classical Charged Particles \/}, 
Addison-Wesley, Reading (1965).
\item{4.} M. Abraham, {\it Theorie der Elektrizit\"at \/}, vol. II,
Springer. Leipzig (1905).
\item{5.} P.A.M. Dirac, Proc. Roy. Soc.(London) {\bf A167}, 148
(1938).
\item{6.} A.A. Sokolov and I.M. Ternov, {\it Radiation from
Relativistic Electrons\/}, Amer. Inst. of
Phys. Translation Series, New York (1986).  
\item{7.} For example, J.M. Aguirrebiria, J. Phys. A:Math. Gen {\bf
30}, 2391), D. Villarroel, Phys. Rev. A {\bf 55}, 3333 (1997),and 
references therein.
\item{8.} F. Rohrlich, Found. Phys. {\bf 26}, 1617 (1996);
Phys. Rev.D{\bf 60}, 084017 (1999); Amer. J. Phys. {\bf 68},1109
(2000).  See also,  H. Levine,
 E.J. Moniz and D.H. Sharp, Amer. Jour. Phys. {\bf 45}, 75 (1977);
A.D. Yaghjian, {\it Relativistic Dynamics of a Charged Sphere},
Springer, Berlin (1992).
\item{9.}  A. Gupta and T. Padmanabhan, Phys. Rev. {\bf D57},7241
(1998).
\item{10.} B. Mashoon, Proc. VII Brazilian School of Cosmology and
Gravitation, Editions Fronti\'eres (1994); Phys. Lett. A {\bf 145},
147 (1990); Phys. Rev. A {\bf 47}, 4498 (1993).
\item{11.} C.W. Misner, K.S. Thorne and J.A. Wheeler, {\it
Gravitation\/}, W.H. Freeman, San Francisco (1973).
\item{12.}  E.C.G.Stueckelberg, Helv. Phys. Acta {\bf 14}, 322(1941); 
{\bf 14}, 588 (1941).
\item{13.} R.P. Feynman, Rev. Mod. Phys. {\bf 20}, 367 (1948).
\item{14.} L.P. Horwitz, Found. Phys. {\bf 25}, 39 (1995).
\item{15.} L. Burakovsky, L.P. Horwitz and W.C. Schieve, Phys. Rev. D
{\bf 54}, 4029 (1996).
\item{16.} D.Saad, L.P.Horwitz and R.I.Arshansky, Found. of Phys. {\bf
19}, 1125 (1989);  M.C. Land, N. Shnerb and L.P. Horwitz,
Jour. Math. Phys. {\bf 36}, 3263 (1995); N. Shnerb and L.P. Horwitz,
Phys. Rev. {\bf 48A}, 4068 (1993).
\item{17.}  O. Oron and L.P. Horwitz, Phys. Lett. {\bf A 280}, 265
(2001).
\item{18.} J.D. Jackson, {\it Classical
Electrodynamics\/},  2nd edition, John Wiley and Sons, New York(1975).
\item{19.} O. Oron and L.P. Horwitz,  ound. Phys. {\bf 33}, 1323
(2003).
\item{20.} J.C. Sprott, {\it Chaos and Time-Series Analysis \/},
Oxford Univ. Press, Oxford (2003).
\item{21.} H.D.I. Abarbanel, {\it Analysis of Observed Chaotic Data\/},
Springer-Verlag, New York (1996); G.L. Baker and J.P. Gollub, {\it
Chaotic dynamics, an introduction\/}, Cambridge Univ. Press, Cambridge (1996).

\vfill
\bye
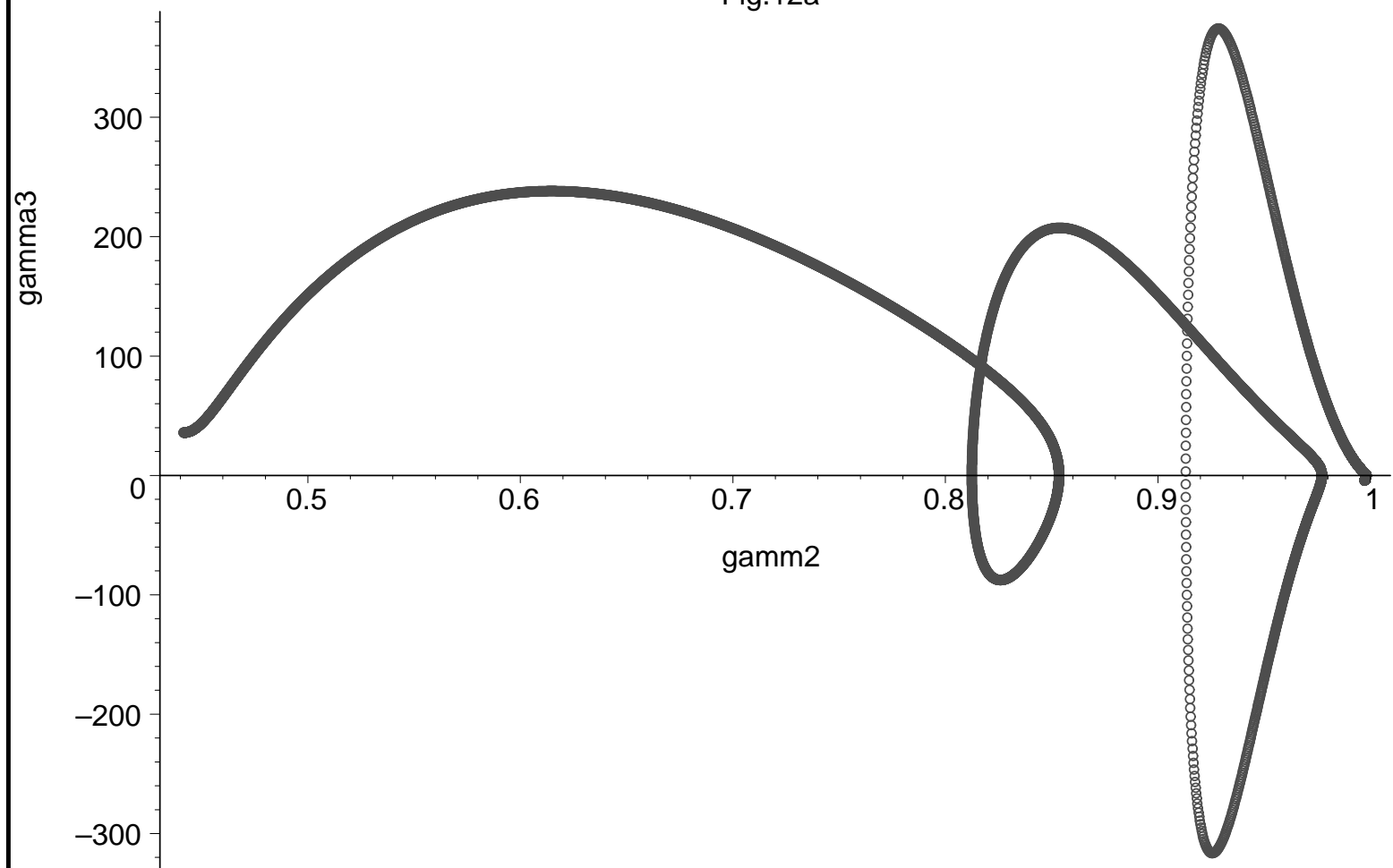
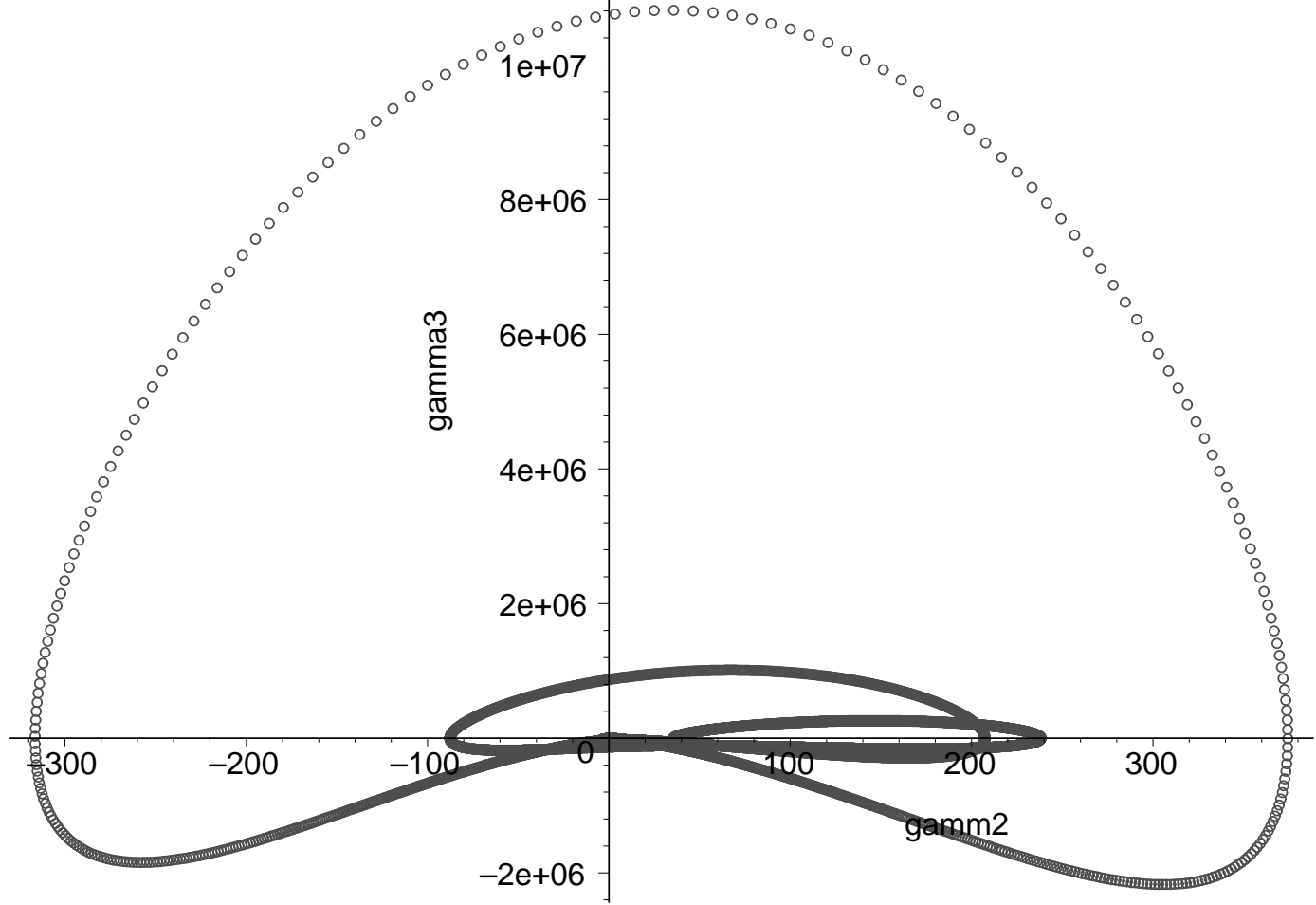
\end